\documentclass[a4paper]{article}

\usepackage{icrc2013}
\usepackage{hyperref}
\usepackage{wasysym}
\usepackage[english]{babel}
\title{The Photodetector Plane of the 4m Davies Cotton Small Size Telescope
for the Cherenkov Telescope Array}

\shorttitle{The PDP of the 4m Davies Cotton SST for CTA}

\authors{
V.~Boccone$^{1}$, 
J.A.~Aguilar$^{1}$, 
A.~Basili$^{1,2}$, 
A.~Christov$^{1}$, 
M.~della Volpe$^{1}$, 
T.~Montaruli$^{1}$ and 
M. Raamez$^{1}$ for the CTA Consortium
}

\afiliations{
$^1$  D\'epartement de physique nucl\'eaire et corpusculaire, Universit\`e de Gen\'eve, CH-1211, Switzerland \\
$^2$  Physik Institut, Universit\"at Z\"urich,  Winterthurerstr. 190, CH-8057, Switzerland 
}

\email{boccone@cern.ch}

\abstract{
Photomultipliers (PMTs) are currently adopted for the photodetector plane of  Imaging Atmospheric Cherenkov Telescopes (IACTs). Even though PMT quantum efficiency has improved impressively in the recent years, one of the main limitation for their application in the gamma-astronomy field -- the impossibility to operate with moon light -- still remains. As a matter of fact, the light excess  would lead to significant and faster camera ageing.  Solid state detectors, in particular Geiger-mode avalanche photo-diodes (G-APDs) represent a valuable alternative solution to overcome this limitation as demonstrated in the field by the FACT experiment (The First G-APD Cherenkov Telescope\cite{FACT:experiment}). They can be regarded as a more promising long term approach, which can be easily adopted for the new generation of cameras and for the Cherenkov Telescope Array (CTA). 
We describe here the Photo-Detector Plane (PDP) of the camera for the 4 m Davies Cotton CTA Small Size Telescopes, for which large area G-APD coupled to non-imaging light concentrators are planned. The PDP includes 1296 photosensors, the biasing and pre-amplification stages, the control electronics as well as the mechanical support and the water-tight enclosure.
We developed with Hamamatsu a new large area hexagonally shaped G-APD with an area of 93.6 mm$^{2}$. This G-APD is divided into 4 channels which will be summed after the pre-amplification stage to maintain an acceptable time characteristic of the signal. The characterization of this device for 50 $\mu$m and 100 $\mu$m micro-cell sizes will be  discussed and compared to other non-custom photodetectors.}


\newcommand{\hide}[1]{}%\textcolor{darkgreen}{\bf \textsc{#1}}
\begin{document}
\maketitle
\section{Introduction}
 The enigma of the origin of cosmic rays and their production mechanics is today, after more than 100 years from their discovery, yet to be solved. Gamma rays and neutrinos can be traced back to their sources, as they are not deflected by the interstellar and intergalactic magnetic fields. The measurement of their spectra from Galactic and extragalactic sources can help us to determine the mechanisms which can lead to the acceleration of particles in cosmic accelerators, such as shock diffusion and jets departing from black holes.
 
 The Cherenkov Telescope Array (CTA) Consortium \cite{Consortium:2010bc} aims at the construction of two large observatories for very high energy gamma rays to cover the full sky. The telescopes are diversified in three different sizes to optimize and extend the sensitivity of the array in various energy ranges from about 20 GeV to 300 TeV. The southern array, profiting from a better exposure to the Galactic plane, will be composed of 4 Large Size Telescopes (LSTs)  to cover the low-energy range up to about $100$~GeV, 24 Middle Size Telescopes (MSTs) for the core energy range from 100 GeV to few TeV and up to 70 Small Size Telescopes (SSTs) to cover the high-energy range above a few TeV to a few hundred TeV. In particular, the LST and SST are discovery instruments since they extend the energy region previously  explored by other experiments respectively in the low and high energy domains. The northern array, smaller in size, will have no SSTs.

During the last year, the FACT collaboration has for the first time demonstrated stable, good performance of a G-APD-based camera for IACTs. It has also demonstrated that data taking with moon light is possible with high photoelectron (pe) threshold values. This is fundamental for the physics goals of SSTs lying in the energy range between \mbox{1-100 TeV}, where longer exposure times are needed to collect the poor photon statistics above 10 TeV.
\section{The Small Size Telescope for CTA}
 There are two sub-projects in CTA for the construction of SSTs: a 4 m-diameter dish Davies-Cotton (DC) single mirror telescope (1M-SST) and a Schwarzschild-Couder type dual mirror telescope (2M-SST) with primary mirror of about 4 m in diameter.  Both the 1M-SST and the 2M-SST propose cameras using G-APDs. The Davies-Cotton design combines a well known structure technology with an innovative camera design based on single photon counting solid state light detectors.

In the past a 7~m design for the DC-SST has been considered adopting PMT-based pixels. The 4 m mirror size implies the usage of smaller devices to compose the pixels (G-APD coupled to non-imaging light concentrators) that translates in a considerable reduction of construction and transportation costs. 
 
In the DC layout mirror facets have radius of curvature $R= 2f$, where $f$ is the focal length of the telescope which corresponds also to the radius of the sphere along which facets lie. This design suffers no global spherical aberrations and provides good imaging over a wide field of view (FoV). The required FoV for SSTs is at least $9^\circ$.
The 1M-SST mirror, described in Ref.~\cite{Rafal, Niemiec}, has an effective diameter of D = 3.98 m. The mirror is composed of 18 hexagonal facets (side-to-side 78 cm with 2 cm separation) and has a focal length of \mbox{$f = 5.6~$m} and a \mbox{FoV = 9$^{\circ}$}. First simulations indicate that the required sensitivity is achieved by an array of 64 telescopes with distance between telescopes larger than 200 m \cite{Rafal}.
 \begin{figure}[t!]
  \centering
  \includegraphics[trim=0cm 0.0cm 0cm 0cm, clip=true,width=0.5\textwidth]{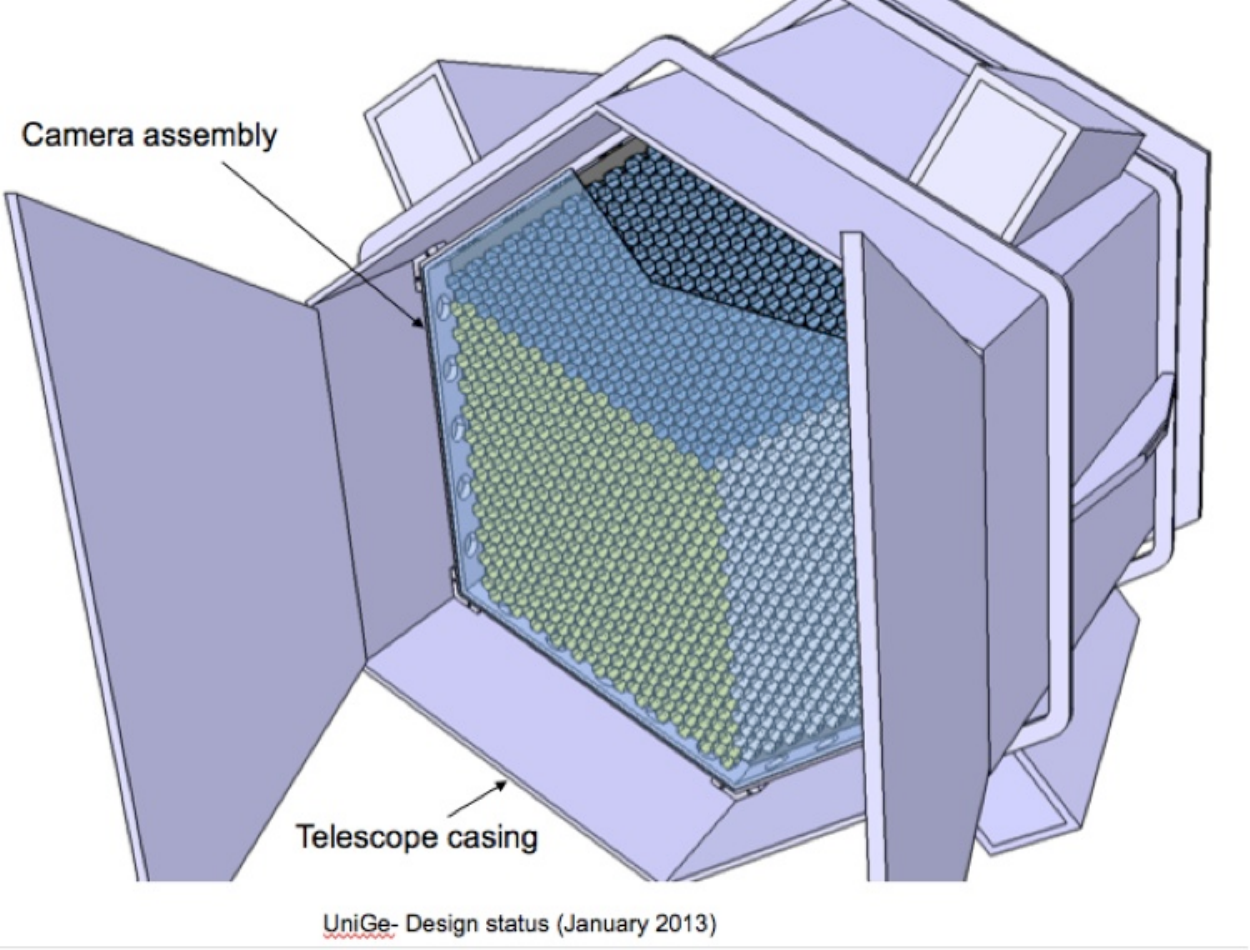}
  \caption{CAD design of the 1M-SST camera}
  \label{fig:1}
 \end{figure}

\section{The camera for the 1M-SST telescope}
\begin{figure}[h!]
\centering
\includegraphics[trim=0cm 8.1cm 0cm 0.1cm, clip=true, width=0.3\textwidth]{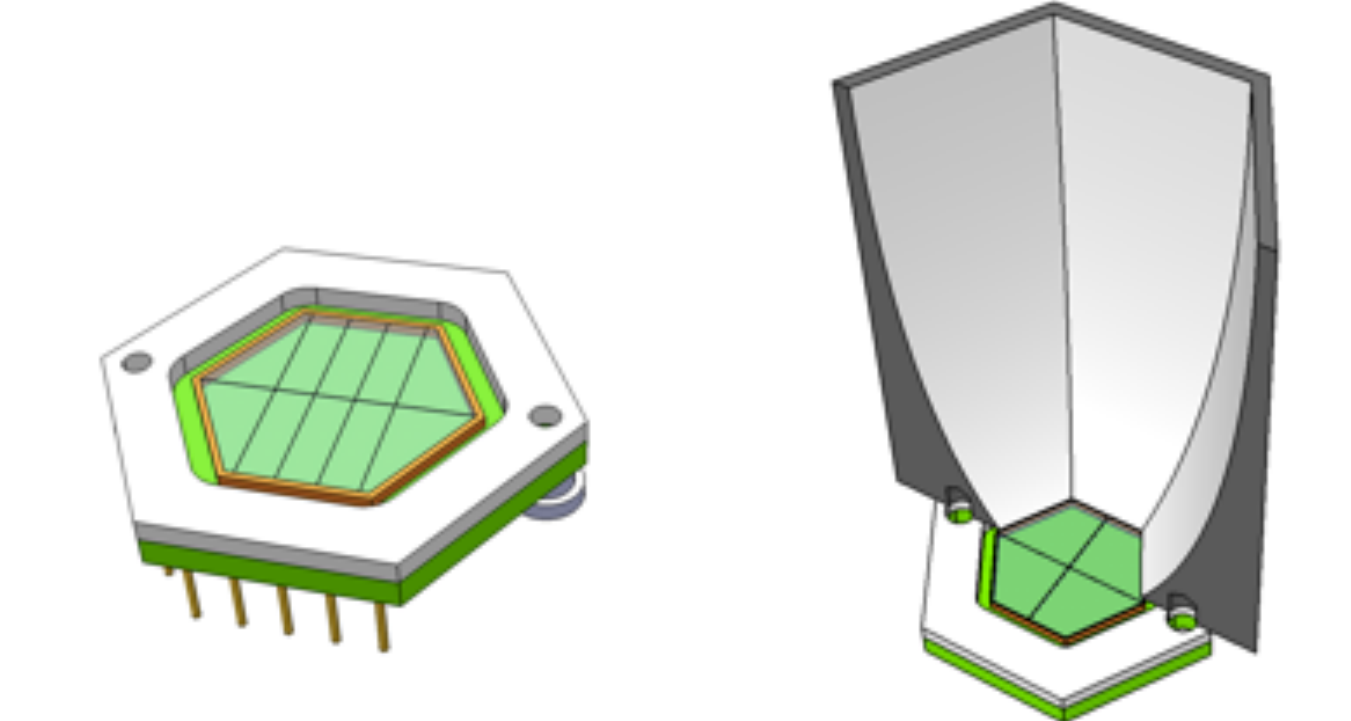}
\caption{Coupling of the large area G-APD to the non imaging light collectors.}
\label{fig:2}
\end{figure}

 The hexagonal camera (shown in Fig.~\ref{fig:2}) has 1296 pixels and the diameter of the circle subtended by the hexagon is 88 cm. The natural linear pixel size $d$ for such a telescope is 
\begin{equation}
 d = f \times p \hide{= \frac{2\pi\cdot 5.6 \cdot 0.25}{180^{\circ}}} \simeq 2.4~\textrm{cm}
\end{equation}
 where $p=0.25^{\circ}$ is the angular pixel size required to achieve the required sensitivity\hide{for the reconstruction of the shower morphology} in the SST. The linear pixel size is smaller than that of the MSTs and LSTs which are made of PMTs coupled to funnels with linear size of about 5 cm. \hide{Hence smaller photodetectors, such as G-APDs, are employed that also have numerous advantages compared to PMTs.} 

 Even though G-APD single channels of 2.4 cm size are not yet available on the market, Hamamatsu agreed upon an R\&D program with UniGE to develop a large enough hexagonal photosensor to match a hollow funnel (see Fig.~\ref{fig:2}). The front-end electronics of the PDP is now being designed while the DAQ/Trigger part of the camera is part of the FlashCam project  \cite{FlashCam}. The camera will be composed of 12-pixel modules that fit the design of the FlashCam for a total of 108 modules. The front-end electronics which includes the pre amplification and driving stage will be embedded in the PDP.  

\section{Characterization of the G-APDs}
The first two large-area G-APD prototypes from Hamamatsu with square micro-cells of 50 and 100 $\mu$m side, respectively called S12516-050 and S12516-100, have been received in December 2012. Both are segmented in 4 channels in order to limit the effect of the capacitance on the signal generation. The S12516-050 has 9210 cells per channel and a fill factor\footnote{The fill factor is defined as the effective active area divided by the total area of the sensor.} of 61.5\% while the S12516-100 has 2282 cells per channel and a fill factor of 78.5\%.

We include in this proceeding the results of the tests for the Hamamatsu S10985-050, which is a $2\times2$ square array of sensors each 3 mm $\times$ 3 mm, and the SenSL $\mu$B30035-X13-E15 that is a newly introduced commercial device with a reduced crosstalk with a microcell pitch of 35 $\mu$m. A list of the devices under test is shown in Tab.~\ref{tab:1}.
\begin{table}[h]
\begin{center}
\begin{tabular}{lcl}\small
Producer and Model        & Channels &Shape/Size\\ \hline
Hamamatsu S10985-050      &  4 &\Square~$6\times6~$mm$^{2}$\\
Hamamatsu S12516-050      &  4 &\hexagon~6~mm side\\ 
Hamamatsu S12516-100      &  4 &\hexagon~6~mm side\\ 
SenSL $\mu$B30035-X13-E15 &  1 &\Square~$3\times3$~mm$^{2}$
\end{tabular}
\caption{Devices under Test}
\label{tab:1}
\end{center}
\end{table}

\subsection{Measurement of the PDE\label{pde:meas}}
The measurement of the PDE was performed using the zero Poissonian method which relies on the simultaneous measurement of the average number of photons  during a triggered light pulse and the average number of photons coming from the background. Both values can be obtained by calculating the ratio between the number of events in which no photons are recorded and the total number of triggers, which makes this value naturally independent of the optical cross-talk effect. The logarithm of this ratio represents the $\mu$ parameter of the Poisson distribution, as shown in Eq.\ref{eqPDE}.
\begin{equation}
\label{eqPDE}\mu = -\log \frac{N_{sig}(0)}{N_{sig}^{tot}} + \log \frac{N_{bkg}(0)}{N_{bkg}^{tot}}
\end{equation}

\begin{figure}[t!]
\centering
\includegraphics[width=0.45\textwidth]{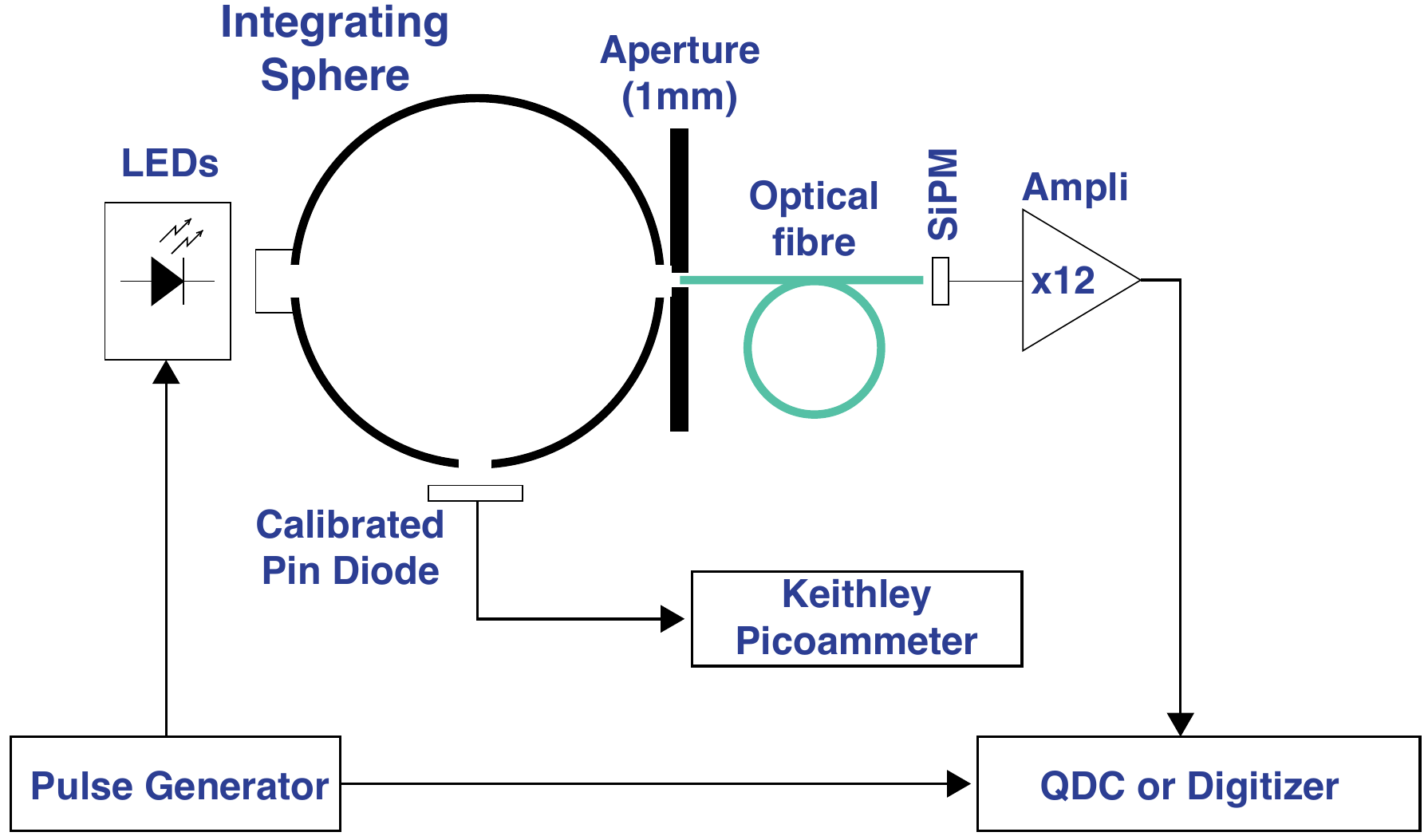}
\caption{\label{fig:3} PDE measurement setup with the integrating sphere.} 
\end{figure}

The light is generated using 7 LEDs at different wavelengths (from 355~nm to 637~nm) pulsed with a $\sim10$~ns width signal. The LEDs are mounted onto a support which is fixed on an open port of a 2" integrating sphere (Thorlab IS200-4).  The light is transported on the detector surface by means of an optical fibre which on one end is connected to the integrating sphere and on the other one is glued onto a metal support which is fixed to the G-APD packaging with two small M2 bolts. 
Each hexagonal G-APD is mounted onto a small PCB mezzanine which can be connected to different amplifier types by standard Micro-MaTch Board to Board connectors. A sketch of the measurement setup is shown in \mbox{Fig.~\ref{fig:3}}.

The calibration of the LED light intensity at the G-APD is obtained by measuring the intensity of the light with a NIST calibrated Hamamatsu photo-diode (S1337-1010BQ) mounted on another port of the sphere behind a 7~mm diameter aperture mask and scaling it with the power ratio between the two light paths defined as:
\begin{equation}
R(\lambda)=\frac{I_{\textrm{meas}}(\lambda)}{I_{\textrm{mon}}(\lambda)}
\end{equation} 
The power ratio must be determined experimentally and has to be such that the current generated by the photo-diode is in a measurable range while only a few photons per pulse arrive to the G-APD surface. Preliminary tests showed that a power ratio of about $5\cdot10^{-3}$ is sufficient to obtain an absolute measurement of the PDE with an error lower than $5-8\%$.  
The power ratio of the sphere is wavelength dependent and needs to be measured during a special calibration where for simplicity the G-APD is replaced by a second calibrated photo-diode. Common mode errors were corrected by repeating the measurement with swapped detectors. The signal of the photo-diode(s) is read out by a Keithley 6487 pico-ammeter connected to a computer. 

 The measurements described here were performed using a general purpose variable gain amplifier board ($12\times$ and $44\times$) based on the OPA695DBV operational amplifier in inverting configuration. The setup is installed inside an aluminium light-tight box  which prevents contamination from the ambient light. The signal of the G-APD are digitized by a LeCroy WaveRunner 610zi oscilloscope sampling with 8 bits at 1~GS/s.% and using a 200 MHz bandwidth filter. 

 The digitized signal is processed offline in order to remove the amplifier offset and to evaluate the charge distribution of the background and of the incoming light. The charge distribution of the light is obtained by integrating the signal in a time window of 100~ns synchronous with the LED pulsed light while the background is obtained instead by integrating the signal in an equivalent time window of about 300~ns prior to the arrival of the LED pulse itself. The repetition rate of the pulser can vary from few kHz to about 50~kHz in order to avoid the effect of the afterpulse. As both charge distributions are obtained by selecting only the first 100~ns of the signal the measured PDE will be independent of the after-pulse effect.

 The data samples are additionally cleaned up by selecting only the events where the baseline was correctly reconstructed. The baseline is obtained by averaging the samples prior to the arrival of the LED pulse. Any dark count signal happening at this time could in fact compromise the correct evaluation of the signal baseline and must be therefore discarded. The probability of those events increases linearly with the dark count rate of the device and is also related to the pulse width of the signal as this determines the total signal occupancy. This selection is performed by removing events with a standard deviation of the baseline greater than $3\sigma$ from the mean.
 
\begin{figure}[h!]
\centering
\includegraphics[width=0.47\textwidth]{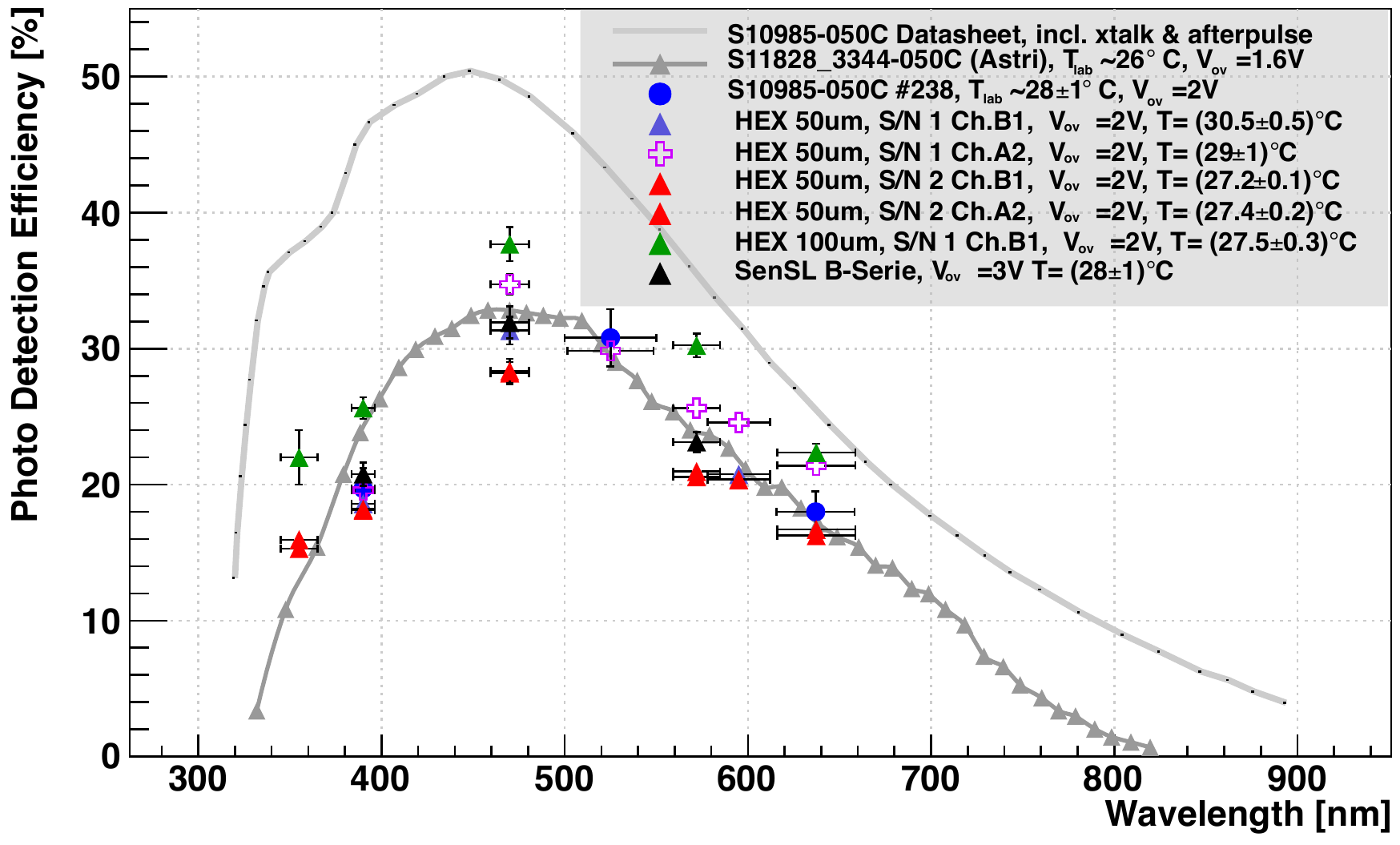}
\caption{\label{fig:4} PDE measurements of the S11828-3344-050C by the ASTRI collaboration (grey triangles), the curve taken from the Hamamatsu data sheet for S10985-050C that includes afterpulses and crosstalk), our measurements of S12516-050 (for 2 channels and 2 operation voltages) and S12516-100 (the two HEX detectors with 50 and 100 $\mu$m cells) and for the SenSL B series.}
\end{figure}

 The pedestal peak of the resulting charge distribution is fit with a gaussian function to determine the integration range ($\mu\pm3\sigma$) used to calculate the number of events corresponding to the $N(0)$ of the Poisson distribution of the signal for the LED pulses and background.
For each wavelength the PDE curve was evaluated as a function of the bias voltage \hide{as shown in Fig.~\ref{figPDE4},} and PDE was evaluated at an over voltage of 2~V, in order to compare later the different detectors on the same working point. The summary of the measured PDE points is shown in Fig.~\ref{fig:4}.

\subsection{Dark count and cross-talk measurement }
For an ideal detector the probability of two or more simultaneous photons thermal excitation should be negligible. For this reason the crosstalk is measured comparing the dark count rate at 0.5 pe with the rate at 1.5 pe.

In order to obtain those values we measure the light pulse rate as a function of the signal pulse height by counting the discriminated signals (in 1 s) for different thresholds. The measurement is performed with a CAEN V814 Low Threshold Discriminator coupled with a CAEN V560 VME scaler. \hide{A C++ program allows automatic operation.} The threshold is increased normally by steps of 1~mV while the step on the biasing voltage depends on the working range of the device under test.
The probability of crosstalk $p_{\textrm{xtalk}}$ will be determined as the ratio:
\begin{equation}
 p_{xtalk}= \frac{f_{>1~pe}}{f_{1~pe}} \approx \frac{f_{>1.5~pe}}{f_{>0.5~pe}}
\end{equation}

Since the typical fall time of the signal is of the order of \mbox{50-80 ns}, a second shaping stage (Ortec NIM shaping amplifier) was added after the preamplifier to reduce the pulse width in order to be able to cope with rates up to \mbox{15-20 MHz}. After the first pre amplification stage -- which is common to the PDE measurement described in Sect.~\ref{pde:meas} -- a shaping stage was added in order to reduce the pulse width to about \mbox{15-20 ns}.
As for the PDE, the cross-talk measurements are performed inside a light-tight black box which prevents contamination from the ambient light, for different G-APD bias voltages (or gain).

\begin{figure}[t]
 \centering
 \includegraphics[width=0.43\textwidth]{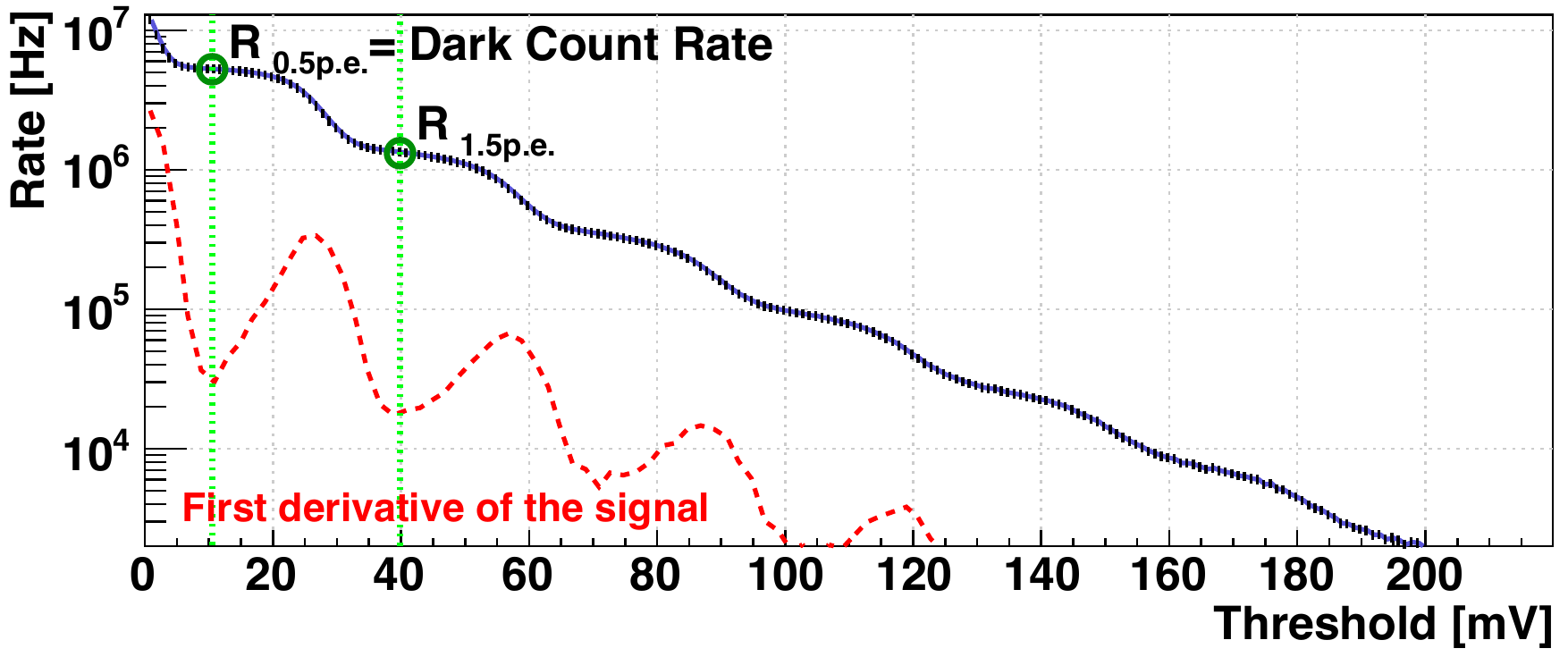}
 \caption{Example of measurement of the dark-count rate and crosstalk.}
 \label{fig:5}\vspace{3mm}
 \centering
 \includegraphics[width=0.43\textwidth]{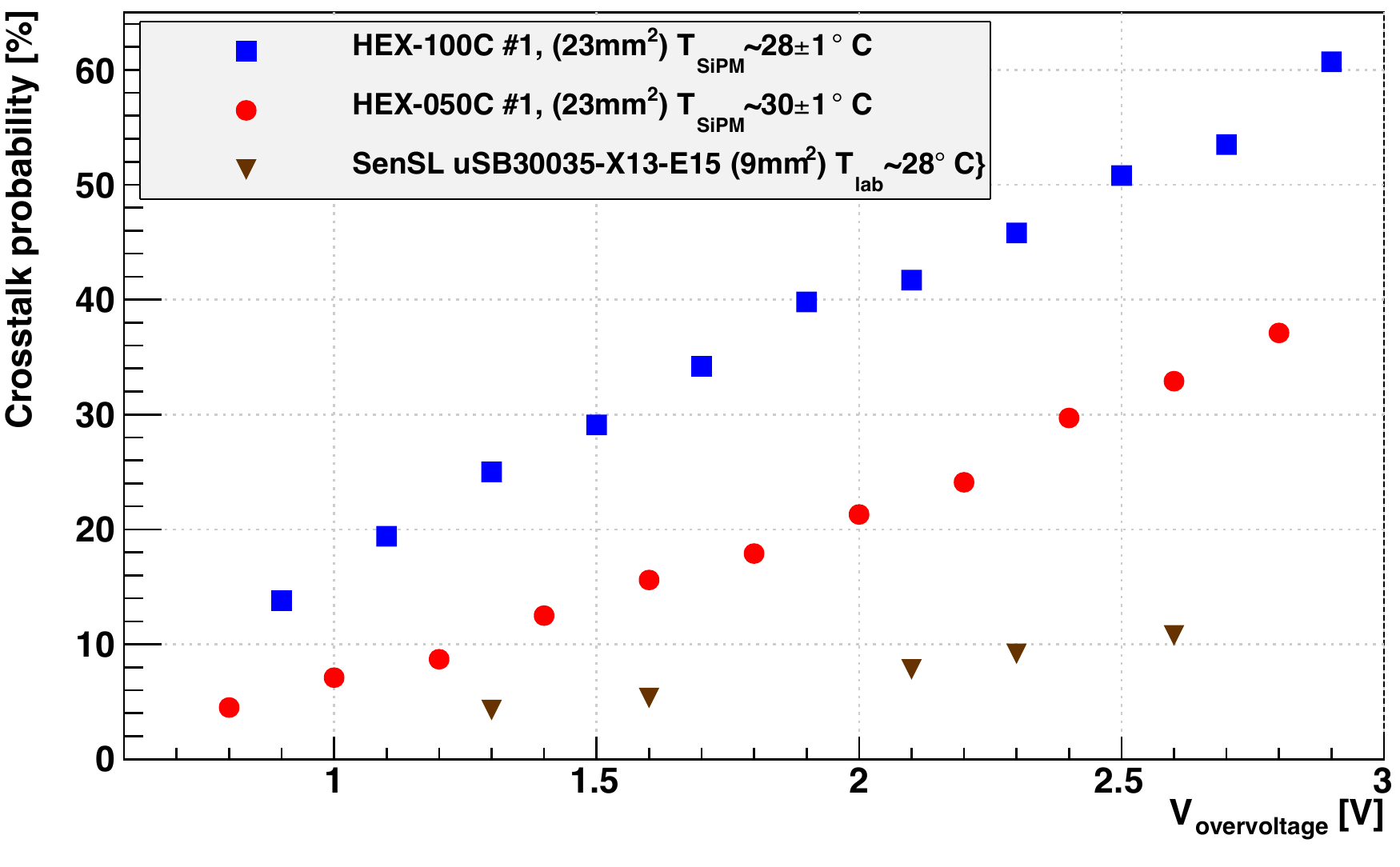}
 \caption{Cross talk probability as a function of the over voltage.}
 \label{fig:6}
\end{figure}

The position of the 0.5~p.e. and 1.5~p.e. points is determined by the valleys in the ${dr}/{dV}$ plot as shown in Fig.~\ref{fig:5} which is obtained by differentiating the measured dark count rate as a function of the threshold voltage (blue line). This measurement was repeated for different bias voltages and the different detectors under test. The summary plot of the cross-talk measurements for the hexagonal sensors is shown in Fig.~\ref{fig:6}\hide{together with the results in \mbox{Ref.~\cite{Eckert}}}.

The 50~$\mu$m Hamamatsu detectors already have a quite high cross talk probability (14\%- 22\%) at bias over voltage ($V_{ov}$)  of 1.6 V which is rather high compared to the modest 5\% of the 50~$\mu$m SenSL. 
The situation changes dramatically when looking at the cross talk probability as a function of $V_{ov}$ as the optimal value of the PDE is reached only for $V_{ov}$ around $2~$V. Under those conditions the 50~$\mu$m detectors have a cross talk probability which is lower than half that measured with the 100~$\mu$m detector.

\begin{figure}[t!]
 \centering
 \includegraphics[width=0.43\textwidth]{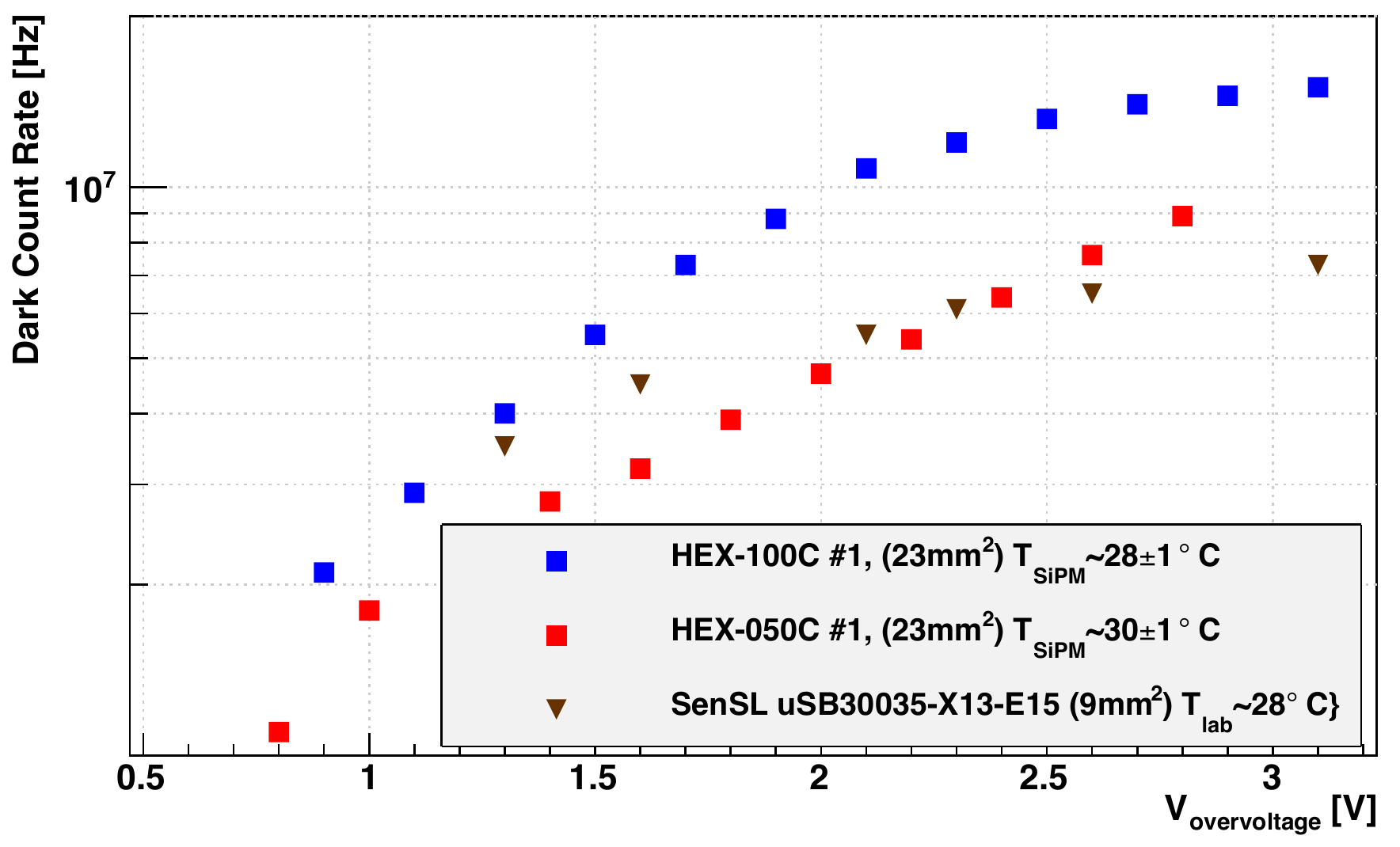}\
 \caption{Absolute dark count rate distribution as a function of the over voltage.}
 \label{fig:7}\vspace{3mm}
 \centering
 \includegraphics[width=0.43\textwidth]{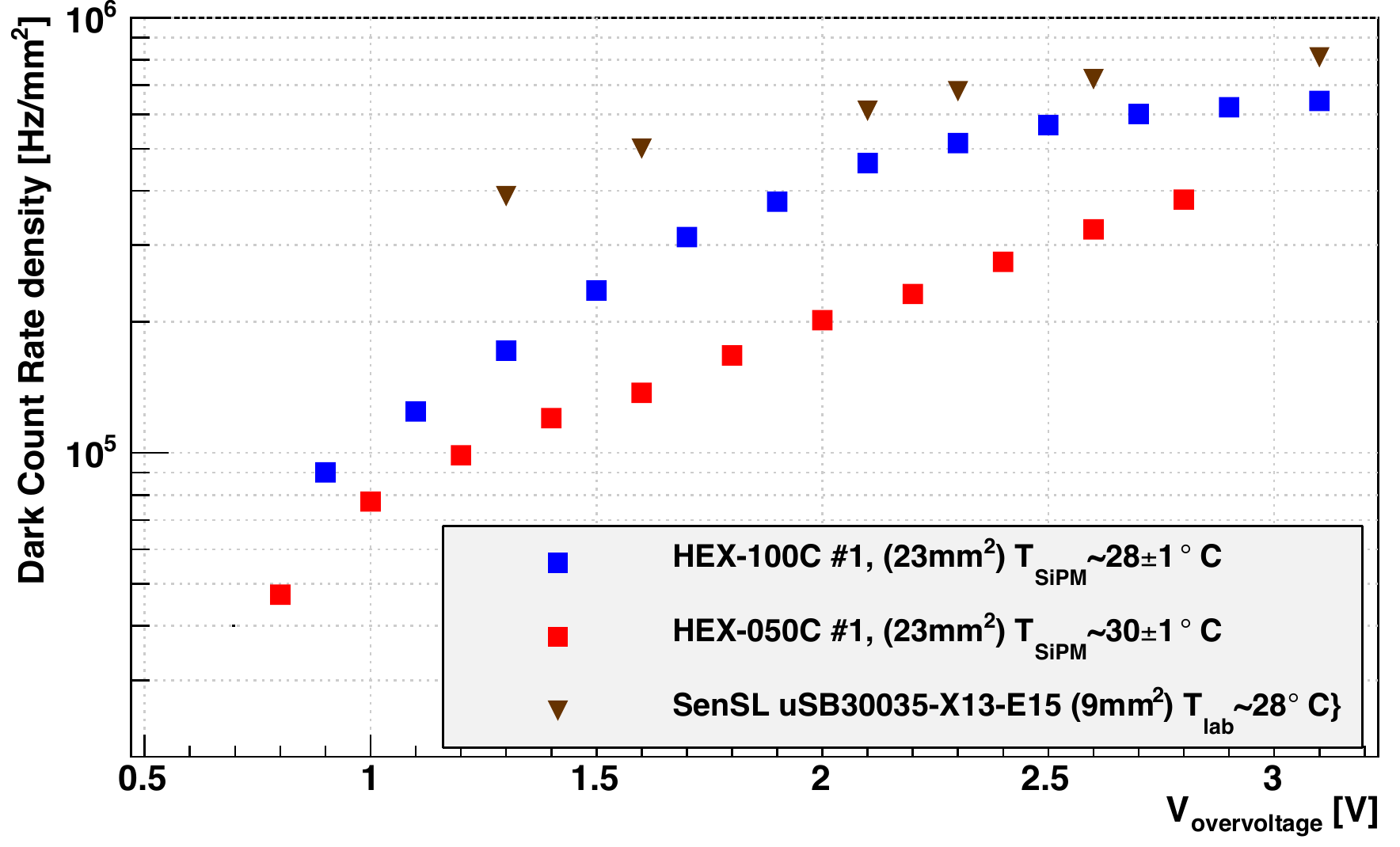}
 \caption{Dark count rate distribution per square mm as a function of the over voltage.}
 \label{fig:8}
\end{figure}

The measurement of the dark count rate is done by taking the rate at 0.5~p.e. ($R_{0.5~p.e.}$) per each individual channel. The results for the two hexagonal G-APDs and the SenSL sensors are shown in Fig.~\ref{fig:7}. The dark count depends on the area therefore it should be normalized to the detector area in order to compare those parameters for different devices. This is shown in Fig.~\ref{fig:8}.

\section{Conclusions}
The first simulations of the array performance using similar properties to the hexagonal photodetectors (Ref.\cite{Rafal}) indicate that the required sensitivity can be achieved. The measurements in the laboratory are showing that the new device has comparable performance to other existing Hamamatsu MPPCs. Moreover, we expect in fall from Hamamatsu new MPPCs with a significant reduction of crosstalk and dark count rate.

We have shown preliminary results for the characterization of G-APDs and described the design of the G-APD based camera for the Davies Cotton SSTs of CTA. We report the first measurement of the characteristics of the new large area hexagonal G-APD which we developed together with Hamamatsu and we compared the results of the PDE, crosstalk and dark count measurements with other commercial devices. The PDE at peak wavelength ($\lambda= 470~$nm) of the S12516-050 hexagonal devices is $(31\pm2)\%$ and is comparable with the S10985-050 and $\mu$B30035-X13-E15. The S12516-100 has a higher PDE $(37.7\pm1.2)\%$, although it has a crosstalk twice as large as its 50~$\mu$m counterpart.

The crosstalk of the SenSL sensor is indeed very low ($<10\%$ at 2~Vov) but the device has a very high dark count rate per unit area (0.6 MHz/mm$^{2}$) which if scaled to the surface area of the hexagon required ($93.6~$mm$^{2}$) brings to an expected dark count rate of about 50~MHz per each device.

\vspace*{0.4cm}
\footnotesize{{\bf Acknowledgment:}{We gratefully acknowledge support from the agencies and organizations   
 listed in this page: \\ \href{http://www.cta-observatory.org/?q=node/22}{http://www.cta-observatory.org/?q=node/22}.}}


\begin{thebibliography}{99}
\bibitem{Consortium:2010bc}
  [The CTA Consortium],
  Exper.\ Astron.\  {\bf 32} (2011) 193
  {\href{http://arxiv.org/abs/1008.3703}{arXiv:1008.3703}} [astro-ph.IM].
\bibitem{FACT:experiment}
 H.~Anderhub {\it et al.} [The FACT Collaboration], 
 {\href{http://arxiv.org/abs/1304.1710}{arXiv:1304.1710}} [astro-ph.IM],
 submitted to JINST (2013). 
\bibitem{Rafal}
 R.~Moderski, {\it et al.}, contrib. \#840, these proceedings, 2013.
\bibitem{Niemiec} 
 J. Niemiec, {\it et al.}, contrib. \#224, these proceedings, 2013.
\bibitem{FlashCam}
 G. Puhlhofer{\it  et al.}, contrib. \#916, these proceedings, 2013. 
\bibitem{Eckert}
 P.~Eckert {\it et al.}
 Nucl.~ Instrum.~Meth.~A {\bf 620}, 2-3, (2010) 217, 
 {\href{http://dx.doi.org/10.1016/j.nima.2010.03.169}{DOI: 10.1016/j.nima.2010.03.169}}
\end{thebibliography}
\end{document}